\pdfoutput=1
\documentclass[acmsmall,nonacm]{acmart}

\acmJournal{TOMACS}

\usepackage{verbatim}
\usepackage{xspace}
\usepackage{amsmath}
\usepackage{multirow}
\usepackage{bm}

\def\..{\,\mathpunct{\ldotp\ldotp}} %
\newcommand{\F}{\mathbf F}

\newcommand{\Z}{\mathbf Z}
\newcommand{\lst}[2]{${#1}_0$,~${#1}_1$, $\dots\,$,~${#1}_{#2-1}$}
\newcommand{\lstm}[2]{{#1}_0$,~${#1}_1$, $\dots\,$,~${#1}_{#2-1}}
\newcommand{\lsto}[2]{${#1}_1$,~${#1}_2$, $\dots\,$,~${#1}_{#2}$}

\newcommand{\xorshift}[1][]{\texttt{xorshift#1}\xspace}
\newcommand{\xoroshiro}[1][]{\texttt{xoroshiro#1}\xspace}
\newcommand{\xoroshirop}[1][]{\texttt{xoroshiro#1+}\xspace}

\newcommand{\xorshiftp}[1][]{\texttt{xorshift#1+}\xspace}

\newcommand{\mt}[1][]{\texttt{MT19937}\xspace}

\begin{document}

\bibliographystyle{ACM-Reference-Format}
\acmJournal{TOMACS}
\title{A New Test for Hamming--Weight Dependencies}
\thanks{This is an updated version of the paper published in \emph{ACM
Transactions on Modeling and Computer Simulation}, 32(3), 2022
(\url{https://doi.org/10.1145/3527582}). With respect to the published version,
Table~\ref{tab:testhwd} reports the word size $w$ of each test, the correct
faulty signature of \xoroshirop[1024], and the missing transitional label of
\xoroshirop[128]; moreover, a new run using the Rust implementation of the test
mentioned below detects bias in WELL512a after $2.5\times 10^{15}$ bytes instead
of $3\times 10^{15}$, we corrected the example batch size for $k=8$ and a few
typos in the text and in the formulas, and we now point to the Rust
implementation instead of the C code}
\author{David Blackman}
\affiliation{%
  \institution{Independent researcher}
  \country{Australia}
}
\author{Sebastiano Vigna}
\orcid{0000-0002-3257-651X}
\affiliation{%
  \institution{Universit\`a degli Studi di Milano}
  \department{Dipartimento di Informatica}
  \country{Italy}
}
\email{vigna@acm.org}

\begin{abstract}
We describe a new statistical test for pseudorandom number generators (PRNGs). Our test
can find bias induced by dependencies among the Hamming weights of the outputs of a PRNG, even
for PRNGs that pass state-of-the-art tests of the same kind from the literature,
and in particular for generators based on $\mathbf F_2$-linear transformations
such as the dSFMT~\cite{MuMPSDPFTNUAT}, \xoroshirop[1024]~\cite{BlVSLPNG},
and WELL512a~\cite{PLMILPGBLRM2}.
\end{abstract}

\ccsdesc[500]{Mathematics of computing~Random number generation}

\keywords{Pseudorandom number generators}

\maketitle

\section{Introduction}

Pseudorandom number generators (PRNGs) are algorithms that generate a seemingly
random output using a deterministic algorithm. A $w$-bit PRNG is defined by a
\emph{state space} $S$, a \emph{transition} (or \emph{next-state}) computable function
$\tau:S\to S$, and a computable \emph{output function} $\varphi:S\to \{\,0,1\,\}^w$ that maps the state space into $w$-bit
words. One then considers an \emph{initial state}, or \emph{seed} $\sigma\in S$, and
computes the sequence of $w$-bit outputs \[ \varphi(\sigma), \varphi(\tau(\sigma)), \varphi\bigl(\tau^2(\sigma)\bigr),
\varphi\bigl(\tau^3(\sigma)\bigr), \ldots \] 
The outputs can be used to generate reals in the unit interval, for example multiplying them by $2^{-w}$.
Knuth discusses PRNGs at length~\cite{KnuACPII}.\footnote{We are here slightly simplifying
	the presentation: in general, the codomain of the output function can be an arbitrary finite set; moreover, depending on the detailed definition, the output
sequence might start with $\varphi(\tau(\sigma))$. }

A classic example is given by
\emph{multiplicative congruential generators}, which are defined by a
prime \emph{modulus} $\mu$ and a \emph{multiplier} $\alpha$. Then $S=\Z/\mu\Z$, $\tau:x\mapsto
\alpha x$, and the output function is given by the binary representation of $x$ (one tries to choose $\mu$ close to $2^w$). Another well-known example is the
class of \emph{$\F_2$-linear generators}~\cite{EcPFLRNG}, in which $S$ is a vector of $\F_2^n$ (i.e., $n$ bits) and $\tau$
is an $\F_2$-linear transformation on $S$; however, usually,
the transformation can be expressed by basic $\F_2$-linear operations on words, such as rotations, shifts, and XORs, rather than in matrix form.
The output function
might pick a word of $w$ bits from the state: for example, if $n=kw$ for some $k$, then
the state can be represented by $k$ $w$-bit words, and the output function
can just choose one of those. In some generators, moreover, the output function is not $\F_2$-linear.

Several theoretical properties help in the design of PRNGs: however, once
designed a PRNG is submitted to a set of \emph{statistical tests}, which
try to discover some statistical bias in the output of the generator. The tests compute, using
the output of the generator, statistics whose distribution is known (at least approximately)
under the assumption that the output of the generator is random. Then, by applying the (complementary) cumulative
distribution function to the statistics one 
obtains a $p$-value, which should be neither too close to zero nor too close to one (see Knuth~\cite{KnuACPII} for a complete introduction to the
statistical testing of PRNGs).

The \emph{Hamming weight} of a $w$-bit word $x$ is the number of ones in its binary representation.
Tests for Hamming-weight dependencies try to discover some statistical bias in
the Hamming weight of the output of the generator.
In particular, such tests do not depend on the numerical
values of the outputs:
indeed, sorting the bits of each examined block (e.g., first all zeroes and then
all ones) would not modify the results of the test (albeit the values would now
be very small).

Since the number of ones in a random $w$-bit word has a binomial
distribution with $w$ trials and probability of success $1/2$, in the most
trivial instance one examines $m$ consecutive outputs \lst xm and checks that
the average of their Hamming weights has the correct distribution (which
will be quickly approximated very well by a normal distribution as $m$
grows~\cite{HiMNASBD}).
Tests may also try to detect \emph{dependencies}: for example, one can
consider (overlapping) \emph{pairs} of consecutive outputs, and check that the
associated pairs of Hamming weights have the expected
distribution~\cite{LESBLCGMF}. Matsumoto and Nishimura~\cite{MaNNTWPNG} have introduced a
theoretical figure of merit that can predict after how many samples
a $\F_2$-linear generator will fail a specific Hamming-weight test. The NIST statistical
test suite for PRNGs~\cite{RSNSTSRPNGCA} contains tests based on Hamming-weight dependencies, too.

In this paper, we introduce a new test for Hamming-weight dependencies that
improves significantly over the state of the art. We find bias in some old
and some new generators for which tests of this type from TestU01~\cite{LESTU01},
a well-known framework for statistical testing of PRNGs,
were unable to find bias even using a large amount of output.

An implementation of our test is available as a Rust
crate,\footnote{\url{https://crates.io/crates/hwd}} which can also examine the
output of a generator in parallel using multiple cores.

\section{Motivation}
\label{sec:mot}

It is known since the early days of $\F_2$-linear generators that sparse
transition matrices induce some form of dependency on the Hamming weight of the
state. Since the output is computed starting from the state, these dependencies
might induce Hamming-weight dependencies in the output, too.
For example, if the state has very low Hamming weight, that is, very few ones,
one might need a few iterations (or more than a million, in the case of the
Mersenne Twister with $19937$ bits of state~\cite{MaNMT}) before the state
contains ones and zeroes approximately in the same amount.
This is however a minor problem because for generators with, say, at least $128$ bits of state,
the probability of passing through such states is negligible.

However, what we witness very clearly in the case of almost-all-zeros state might be true in general:
states with few ones might lead to states with few ones, or due to XOR operations states with many
ones might lead to states with few ones. This kind of dependency is more difficult to detect.

Here we consider as motivation a few generators:
\xorshiftp[128]~\cite{VigFSMXG} is the stock generator of most JavaScript implementations in 
common browsers; the SFMT
(SIMD-oriented Fast Mersenne Twister)~\cite{SaMSOFMT} is a recent improvement on the classic Mersenne
Twister using SIMD instructions, and we will use the version with $607$ bits of state; the dSFMT~\cite{MuMPSDPFTNUAT} 
is another version of the Mersenne Twister which natively generates doubles, and we will use the version with $521$ bits of state; 
WELL is a family of generators with excellent equidistribution properties~\cite{PLMILPGBLRM2}, and
we will use the version with $512$ bits of state;
finally,
we consider a new $\F_2$-linear transformation, \xoroshiro, designed by the authors, and the
associated generators \xoroshirop[128] and \xoroshirop[1024]~\cite{BlVSLPNG}.\footnote{The generators
combine two words of state using a sum in $\Z/2^{64}\Z$.}
All these generators have quite sparse transition matrices (WELL512a having the densest matrix), and one would expect some kind
of Hamming-weight dependency to appear.

To check whether this is true, we can test for such dependencies using
TestU01~\cite{LESTU01}, a well-known framework for testing generators, which
implements tests related to Hamming weights from~\cite{LESBLCGMF} and~\cite{RSNSTSRPNGCA} (please refer to
the TestU01 guide for a detailed description of the tests). Table~\ref{tab:params} shows the basic parameters
of the nine tests we performed. The parameters were inspired by the authors' choices in the
BigCrush test suite~\cite{LESTU01}, but instead of analyzing $10^{9}$ or fewer outputs, as happens in BigCrush,
we analyze up to $10^{13}$ $32$-bit values (e.g., $40$\,TB of data), hoping that
examining a much larger amount of data might help in finding bias in the
generators. Besides the parameter inspired by BigCrush, following a suggestion from a referee 
we also tried a number of power-of-two values for the $L$ parameter of HammingIndep and HammingCorr.

Some of the generators have a $64$-bit output, but TestU01 expects generators to
return $32$-bit integers, so we tested both the lower $32$ bits, the upper
$32$ bits, and the upper and lower bits interleaved. (We do not discard any bits, as it is possible in TestU01.)
In the case of the dSFMT, there is a specific call to generate a $32$-bit value.

The disappointing results are reported in Table~\ref{tab:TestU01}: despite the very
sparse nature of the transition matrices of these generators, and the very large
amount of data, only problems with the SFMT (already using $10^9$ values), \xorshiftp[128] (using $10^{10}$ values),
and \xoroshirop[128] (using $10^{13}$ values)
are reported.\footnote{The latter failure emerged only after testing with additional values of the parameter $L$
suggested by one of the referees.} All other $p$-values at $10^{13}$ are within the range $[0.01\..0.99]$.\footnote{In two cases
we found $p$-values slightly outside this range, but a test using $1.2\times 10^{13}$ values showed that they
were statistical flukes.}

\begin{table}
\caption{\label{tab:params}Parameters for statistical tests related to Hamming weights 
from TestU01~\cite{LESTU01}. We consider the entire output of the generator (i.e., TestU01 parameters $r=0$, $s=32$). The parameter
$d$ of HammingIndep has been set to zero. The parameter $k$ varies among $30$, $300$, and $1200$ for HI and $30$, $300$, $500$
for HC, as in BigCrush. Moreover, in both cases we tested $k$ equal to $128$, $256$, $512$, and $1024$ following a suggestion from a referee.}
\renewcommand{\arraystretch}{1.1}
\begin{tabular}{lr}
\multicolumn{1}{c}{Label} & \multicolumn{1}{c}{Test parameters}\\
\hline  
HW0 & HammingWeight2 ($N=1$, $L=10^6$) \\
HW1 & HammingWeight2 ($N=1$, $L=10^7$) \\
HW2 & HammingWeight2 ($N=1$, $L=10^8$) \\
HI$k$ & HammingIndep ($N=1$, $L=k$) \\
HC$k$ & HammingCorr ($N=1$, $L=k$) \\
\end{tabular}
\end{table}

\begin{table}
\caption{\label{tab:TestU01}Results for the TestU01~\cite{LESTU01} statistical tests related to Hamming weights.
The $n$ parameter gives the number of $32$-bit outputs examined. The tests have been run on the lower $32$ bits of the output (``L''), 
the upper $32$ bits (``U''), and interleaving the upper and lower bits (``I''). We report the first test failed by a generator,
where failure is a $p$-value outside of the range $[0.01\..0.99]$.}
\renewcommand{\arraystretch}{1.1}
\begin{tabular}{lrrrrr}
\multicolumn{1}{c}{$n$} & \multicolumn{1}{c}{$n=10^{9}$}  & \multicolumn{1}{c}{$n=10^{10}$}  & \multicolumn{1}{c}{$n=10^{11}$}  & \multicolumn{1}{c}{$n=10^{12}$}  & \multicolumn{1}{c}{$n=10^{13}$} \\
\hline  
SFMT ($607$ bits) & HI512 (I) &\\
\xorshiftp[128] & \multicolumn{1}{c}{---} & HI128 (U) \\
dSFMT ($521$ bits) & \multicolumn{1}{c}{---} & \multicolumn{1}{c}{---} & \multicolumn{1}{c}{---} & \multicolumn{1}{c}{---} & \multicolumn{1}{c}{---}\\
WELL512a & \multicolumn{1}{c}{---} & \multicolumn{1}{c}{---} & \multicolumn{1}{c}{---} & \multicolumn{1}{c}{---} & \multicolumn{1}{c}{---}\\
\xoroshirop[128] & \multicolumn{1}{c}{---} & \multicolumn{1}{c}{---} & \multicolumn{1}{c}{---} & \multicolumn{1}{c}{---} & HI128 (I)\\
\xoroshirop[1024] & \multicolumn{1}{c}{---} & \multicolumn{1}{c}{---} & \multicolumn{1}{c}{---} & \multicolumn{1}{c}{---} & \multicolumn{1}{c}{---} \\
\end{tabular}
\end{table}

\section{Testing Hamming-weight dependencies}
\label{sec:hwd}

In this section, we introduce a new test for Hamming-weight dependencies that
will find bias in all generators from Table~\ref{tab:TestU01}. For the
generators whose bias was detected by TestU01, the test will be able to obtain
similar or better $p$-values using an order of magnitude fewer data.

Let us denote with $\nu x$~\cite{KnuACPIV} the Hamming weight of a $w$-bit word $x$, that
is, the number of ones in its binary representation. We would like to examine the output of
a generator and find \emph{medium-range} dependencies in the Hamming weights of its outputs.

For example, the current output might have an average weight (close to $w/2$)
with higher-than-expected probability if three outputs ago we have seen a word
with average weight; or, the current output might have a non-average weight
(high or low) with higher-than-expected probability depending on whether four
outputs ago we have seen average weight \emph{and} five outputs ago we have seen
non-average weight.

We will consider sequences of $w$-bit values ($w$ even and not too small, say
$\geq 16$) extracted from the output of the generator. Usually, $w$
will be the entire output of the generator, but it is possible to run the test on a
subset of bits, break the generator output into smaller pieces fed
sequentially to the test, or glue (a subset of) the generator output into larger
pieces.

The basic idea of the test is that of generating a vector whose coordinates
should appear to be drawn from independent random variables with a standard normal
distribution, given that the original sequence was random; apply a unitary
transformation, obtaining a transformed vector; and derive a $p$-value using the
fact that the coordinates of the transformed vector should still appear to be
drawn from independent random variables with a standard normal
distribution~\cite{TonMND}, given that the original sequence was random. The transform
will be designed in such a way to make dependencies as those we described emerge more clearly.

First of all, we must define a \emph{window} we will be working on:
thus, we fix a parameter $k$ and consider overlapping $k$-tuples of consecutive
$w$-bit values (ideally, the number of bits of state should be less than $kw$).
We will write $\langle x_0, x_1, \ldots, x_{k-1}\rangle$ for such a generic
$k$-tuple.

Now we need to classify outputs as ``average'' or ``extremal''
with respect to their Hamming weight.
We thus consider an integer parameter $\ell\leq w/2$, and the map 
\[
x\stackrel{d}{\mapsto}\begin{cases}
0& \text{if $\nu x<w/2 -\ell$;}\\
1& \text{if $w/2 -\ell\leq \nu x\leq w/2 + \ell$;}\\
2& \text{if $\nu x> w/2 + \ell$.}
\end{cases}
\]
In other words, we compute the Hamming weight of $x$ and categorize $x$
in three classes: left tail (before the $2\ell+1$ most frequent
weights), central (the $2\ell+1$ central, most frequent weights), right tail (after the $2\ell+1$ most
frequent weights). The standard choice for $\ell$ is the integer such
that the overall probability of the $2\ell+1$ most frequent weights is closest to $1/2$.
For example, for $w=32$ we have $\ell=1$, whereas for $w=64$ we have $\ell=2$.

We thus get from the $k$-tuple $\langle x_0, x_1, \ldots, x_{k-1}\rangle$ a \emph{signature} $\langle d(x_0), d(x_1), \ldots, d(x_{k-1})\rangle$ of $k$
\emph{trits} (base-$3$ digits), which we will identify with its value as a number in base $3$:
\[
\sum_{i=0}^{k-1}d(x_i)3^{k-1-i}.
\]

Now, given a sequence of $m$ $w$-bit values, for each signature $s$ we
compute the average number of ones in the word appearing after a $k$-tuple
with signature $s$ in the sequence. More precisely, a subsequence of the form
$\langle x_0, x_1, \ldots, x_k\rangle$ contributes $\nu x_k$
to the average associated with the signature $\langle d(x_0), d(x_1), \ldots, d(x_{k-1})\rangle$.

 This bookkeeping can be easily performed
using $3^k$ integer variables while streaming the generator output.
For a large $m$, this procedure yields $3^k$ values with
approximately normal distribution~\cite{HiMNASBD},\footnote{In practice, one must size $m$ depending on the 
number of signatures, so that each signature has a sufficiently large number of associated samples. Implementations
can provide quickly the user with a preview output  
using the value zero for random variables associated with non-appearing signatures, 
warning the user that final $p$-values too close to one might be artifacts.} which we normalize to a
standard normal distribution; we denote the resulting row vector with $\bm v=\langle \lstm v{3^k}\rangle$.\footnote{Breaking the generator output in smaller pieces provides
obviously a finer analysis of the distribution of each piece, but a test with
parameters $k$ and $w$ ``covers'' $kw$ bits of output: 
if we analyze instead values made of $w/2$ bits, to
cover $kw$ bits of output we need to increase the length of tuples to
$2k$, with a quadratic increase of memory usage.}\footnote{There is also a
\emph{transitional} variant of the test: we see the sequence of $w$-bit values as a stream of
bits, xor the stream with itself shifted forward by one bit, and run the test on
the resulting $w$-bit values. In practice, we look for Hamming-weight
dependencies between bit \emph{transitions}.}

We now apply to $\bm v$ a Walsh--Hadamard-like transform, multiplying $\bm v$ 
by the $k$-th Kronecker power\footnote{For a definition of the Kronecker product, see~\cite[Section~4.3]{ZhaMT}.} $T_k$ of the unitary base matrix
\begin{equation}
\label{eq:transform}
M=\left(\begin{matrix}
\frac1{\sqrt3} & \phantom{-}\frac1{\sqrt2} & \phantom{-}\frac1{\sqrt6}\\
\frac1{\sqrt3} & \phantom{-}0 & -\frac2{\sqrt6}\\
\frac1{\sqrt3} & -\frac1{\sqrt2} & \phantom{-}\frac1{\sqrt6}\\
\end{matrix}\right).
\end{equation}
Assuming that $T_k$ is indexed using sequences of trits as
numerals in base $3$, the transform can be implemented recursively in the same
vein as the fast Walsh--Hadamard transform (or any transform based on Kronecker powers), as
if we write $\bm v= \bigl[\bm v^0\;\bm v^1\;\bm v^2\bigr]$, where $\bm v^0$, $\bm v^1$, and $\bm v^2$ are the three subvectors
indexed by signatures
starting with $0$, $1$, and $2$, respectively, we have by definition $T_0=1$ and
\begin{multline}
\label{eq:rec}
\bm v T_k=\bigl[\bm v^0\;\bm v^1\;\bm v^2\bigr]\left(\begin{matrix}
\frac1{\sqrt3}T_{k-1} & \phantom{-}\frac1{\sqrt2}T_{k-1} & \phantom{-}\frac1{\sqrt6}T_{k-1}\\
\frac1{\sqrt3}T_{k-1} & \phantom{--}0 & -\frac2{\sqrt6}T_{k-1}\\
\frac1{\sqrt3}T_{k-1} & -\frac1{\sqrt2}T_{k-1} & \phantom{-}\frac1{\sqrt6}T_{k-1}\\
\end{matrix}\right)\\
=\left[\frac1{\sqrt3}\left(\bm v^0 +\bm v^1 +\bm v^2\right)T_{k-1} \quad
\frac1{\sqrt2}\left(\bm v^0 -\bm v^2\right)T_{k-1}  \quad
\frac1{\sqrt6}\left(\bm v^0 -2\bm v^1 +\bm v^2 \right)T_{k-1} \right].
\end{multline}
A detailed C implementation of $T_k$ will be described in Section~\ref{sec:transform}.

We will denote the transformed vector by $\bm v'= \bm v T_k$, and
we shall write $v'_i$ for the transformed values.
Since $T_k$ is unitary, the $v'_i$'s must
appear still to be drawn from a standard normal distribution, and 
we can thus compute $p$-values for each of them.
We combine $p$-values by dividing the indices of the vector $\bm v'$ 
in $C$ categories \lsto{\mathscr C}C using the number of nonzero trits contained in their base-$3$ representation,
that is, the number of nonzero trits in the associated signature: 
$\mathscr C_j$, $1\leq j<C$, contains indices with $j$ nonzero trits,
whereas $\mathscr C_C$ contains all remaining indices, whose base-$3$ representation
has at least $C$ nonzero trits ($C\leq k$; usually,
$C=\lfloor k/2\rfloor + 1$). We discard $v'_0$.

Given a category, say of cardinality $c$, we have thus a $p$-value $p_i$ for each $v'_i$ in the category; 
we then consider the minimum of the $p$-values, say, $\bar p$, and compute a final category $p$-value
by composing with the cumulative distribution function
of the minimum of $c$ independent uniform random variables in the unit interval~\cite[Eq. (2.2.2)]{DaNOS},
obtaining the category $p$-value $1-(1-\bar p)^c$. Finally, we take the minimum category $p$-value
over all categories, and apply again the same cumulative distribution function with parameter $C$, since we are taking
the minimum over $C$ categories: this yields the final $p$-value of the test. Formally,
\[
p = 1 - \left( 1 -  \min_{1\leq j \leq C}\left(1 - \left( 1 -  \min_{i\in \mathscr C_j}  p_i    \right)^{\left|\mathscr C_j\right|}\right)\right)^C.
\]

The point of the transform $T_k$ is that while $v_i$ represents the (normalized) average number of
ones after $k$ previous outputs with density pattern described by the trit representation of $i$, 
$v'_i$ represents a combination of the average number of
ones after $k$ previous outputs satisfying different constraints: in the end, a unitary
transformation is just a change of coordinates.

As a simple example, let us write the index $\bar\imath$ of a transformed value
$v'_{\bar\imath}$ as a sequence of trits \lsto tk. If the trits are all zero,
looking at~(\ref{eq:rec}) one can see that we are just
computing the normalized sum of all values, which is of little interest: indeed, we discard $v'_0$.

On the contrary, if a single trit, say in position $\bar\jmath$, is equal to
$1$, $v'_{\bar\imath}$ is given by the sum of all $v_i$'s in which the
$\bar\jmath$-th trit of $i$ is $0$ ($\bar\jmath$ steps before we have seen few ones) minus the sum of all $v_i$'s in
which the $\bar\jmath$-th trit of $i$ is $2$ ($\bar\jmath$ steps before we have seen many ones): if the
Hamming weight of the output depends on the Hamming weight of the output 
$\bar\jmath$ steps before, the value of $v'_{\bar\imath}$ will be
biased. Intuitively, if the transition matrix is very sparse we expect vectors with
low or high Hamming weight to be mapped to vectors with the same property.

If instead a single trit in position $\bar\jmath$ is equal to $2$ we will
detect a kind of bias in which the Hamming weight of the current value depends on
whether the Hamming weight of the output $\bar\jmath$ steps before was 
extremal or average: more precisely, whether the value is larger
when the Hamming weight of the output $\bar\jmath$ steps before was average and smaller
when the Hamming weight of the output $\bar\jmath$ steps before was extremal (and \textit{vice versa}).
Intuitively, we expect that the shift/rotate-XOR of states with a very small or very large number of ones
will have a small number of ones (in the first case, by sparsity, in the second case, by cancellation).

More complex trit patterns detect more complex dependencies:  the most
interesting patterns, however, usually are those with few nonzero trits, as a zero trit
acts as a ``don't care about that previous output'': this property is immediate
from~(\ref{eq:rec}), as the first $3^{k-1}$ output values, which correspond
to a ``don't care'' value in the first position, are obtained by applying
recursively $T_{k-1}$ over the renormalized sum of $\bm v_0$, $\bm v_1$, and $\bm v_2$, thus 
combining the values associated with signatures identical but for the first trit.

This is also why we assemble $p$-values by categories: by putting
indices with a higher chance of giving low $p$-values in small categories, the
test becomes more sensitive.

\subsection{Results}

We ran tests with $w=32$ or $w=64$ and $k$ ranging
from $8$ to $19$, depending on the state size.
We performed the tests incrementally, that is, for increasingly larger values of
$m$, and stopped after a petabyte ($10^{15}$ bytes) of data or if we detected
a $p$-value smaller than $10^{-20}$.

Table~\ref{tab:testhwd} reports some generators failing our test.
All generators considered other than \xorshift pass BigCrush, except for
\emph{linearity tests}~\cite{MarTMSRNS,CarALC,ErdETBK} (called MatrixRank and LinearComp in TestU01).
We report the faulty signature, that is, the pattern of dependencies that caused
the low $p$-value: it provides interesting insights into the structure of the
generator. Indeed, we can see that for generators that cycle through their state
array, combining a small part of the state, the test can locate exactly
the dependencies from those parts:
for example, the Hamming weight of the output of \xorshift[1024] depends, not
surprisingly, on the Hamming weight of the first and last word of state.

First, we examine a \xorshift generator~\cite{MarXR} with $128$ bits of state,
and its variant \xorshiftp[128] that we discussed in Section~\ref{sec:mot}. We
can find bias in the latter using just $6$\,GB of data.
Analogously, we find bias in a \xorshift generator using $1024$ bits of state,
and in the SFMT~\cite{SaMSOFMT} with $607$ bits of state
using just $400$\,MB of data. On these extremely simple generators, the performance of
the test is thus in line with that of the tests in TestU01.

However, once we turn to the other generators in Table~\ref{tab:TestU01} the
situation is different: we can find bias in all generators, sometimes using an order of
magnitude less data than in Table~\ref{tab:TestU01}.

Our test can also find Hamming-weight dependencies in some generators of the
Mersenne Twister family with small-to-medium size. First of all, we consider the $64$-bit Tiny Mersenne Twister~\cite{SaMTMT}, which
has $127$ bits of state and a significantly more complex
structure than the other generators in Table~\ref{tab:testhwd}. Moreover, unlike other members of the Mersenne Twister family,
the output function of the Tiny Mersenne Twister contains a non-$\F_2$-linear operation---a sum 
in $\Z/2^{64}\Z$.
To find the bias, we had to resort to a slightly more detailed analysis, 
using $w=32$ and breaking up the $64$-bit output of the generator into two
$32$-bit words. We report a range of results because we tried a few parameters
published by the authors.

We also analyzed the classic Mersenne Twister~\cite{MaNMT} at $521$ and $607$ bits.
We used Matsumoto and Nishimura's library for the dynamic creation of Mersenne Twisters~\cite{MaNDCPNG},
and generated eight different instances of each generator: this is why we report in Table~\ref{tab:testhwd} a range of
values and multiple signatures.
The $607$-bit version performs much worse than the $521$-bit version (in fact, all instances we tested failed
even the classical Gap test from BigCrush). But, more importantly, we
found huge variability in the test results depending on the parameter generated by the library:
in some cases, the $607$-bit Mersenne Twister performs in our test similarly to a \xorshift[128] generator, which has a simpler structure
and a much smaller state. 

Finally, we were able to find bias in
WELL512a~\cite{PLMILPGBLRM2}. In this case, we noticed that the $p$-value was slowly drifting
towards zero at about 1\,PB of data, so we continued the test until it passed the threshold $10^{-20}$.

A comparison between Table~\ref{tab:TestU01} and Table~\ref{tab:testhwd} shows clearly that our
new test is significantly more powerful than the tests of the same kind available in TestU01, as it can detect bias
on $\F_2$-linear generators for which no such bias was previously detectable. In fact, to the best of our knowledge
this is the first time that Tiny Mersenne Twister, the dSFMT at $521$ bits, and WELL512a fail a test
that is not a linearity test.

It is worth noting that \xoroshirop[128] fails in Table~\ref{tab:TestU01} only
when the parameter $L$ is set to $128$, which yields a low $p$-value ($\approx 3\times 10^{-11}$). Larger
values of $L$ (e.g., $256$, $300$,\ldots) do not yield a failure. This is in sharp contrast with our test, where testing
with $k'>k$ will preserve the failures found in dimension $k$, because if $s$ is a $k$-dimensional failing signature,
then $0^{k'-k}s$ will be a failing $k'$-dimensional signature, with some small adjustments due to
the different scaling to the standard normal distribution and the different size of categories. In other words, increasing the dimension of the
test will not prevent the test from detecting bias that was previously detectable at a lower dimension:
the same does not happen for the HammingIndep test of TestU01.

\begin{table}\caption{\label{tab:testhwd}Detailed results of the test described
in Section~\ref{sec:hwd}. We report the word size $w$ and the number of bytes generating a
$p$-value smaller than $10^{-20}$. We report also the trit signature which
caused the low $p$-value. Ranges (represented using the arrow symbol $\rightarrow$) appear when we tried several variants: a missing
right extreme means that some instances did not fail the test within the $1$\,PB limit.}
\renewcommand{\arraystretch}{1.1}
\begin{tabular}{lrrl}
Generator & \multicolumn{1}{c}{$w$} & \multicolumn{1}{c}{$p=10^{-20}$ @} & Faulty signature \\
\hline
\xorshift[128]  & $64$ & $8\times 10^8$ & 00000021 \\
\xorshiftp[128]  & $64$ &$6\times 10^9$ & 00000012 (transitional) \\
\xorshift[1024] & $64$ & $6\times 10^8$ & 2000000000000001 \\
\xorshiftp[1024] & $64$ & $9\times 10^9$ & 2000000000000001 (transitional)\\
\xoroshiro[128]  & $64$ & $1\times 10^{10}$ & 00000012 \\
\xoroshirop[128]  & $64$ &$5\times 10^{12}$ & 00000012 (transitional) \\
\xoroshiro[1024] & $64$ & $5\times 10^{12}$ & 1100000000000001 \\
\xoroshirop[1024] & $64$ & $4\times 10^{13}$ & 1200000000000001 (transitional)\\
Tiny Mersenne Twister ($127$ bits) & $32$ & $8 \times 10^{13}\rightarrow$ & 00000202 \\
\multirow{2}{*}{Mersenne Twister ($521$ bits)} & \multirow{2}{*}{$32$} & \multirow{2}{*}{$4\times 10^{10}\rightarrow$} & 1000000100000000 \\
&&& 2000000100000000\\
\multirow{2}{*}{Mersenne Twister ($607$ bits)} & \multirow{2}{*}{$32$} & \multirow{2}{*}{$4\times 10^{8}\rightarrow 4\times10^{10}$}&  1000000001000000000\\
&&&2000000001000000000 \\
SFMT ($607$ bits) & $64$ & $4\times 10^{8}$&001000001000 \\
dSFMT ($521$ bits) & $32$ & $6 \times 10^{12}$ &1001000100100010\\
WELL512a & $32$ & $2.5 \times 10^{15}$ &2001002200000000\\
\end{tabular}
\end{table}

\section{Implementation details}

We will now discuss some implementation details.
To be able to perform our test in the petabyte range, it must be engineered carefully:
in particular, the main loop enumerating the
output of the generator and computing the values $v_i$ must be as fast as possible.
Counting the number of ones in a word can be performed using single-clock
specialized instructions in modern CPUs. 
The $v_i$'s are stored in an array of $3^k$ elements indexed by the value 
of a signature as a numeral in base $3$, as required by the recursive implementation of $T_k$ (see Section~\ref{sec:transform}). 
One can keep track very
easily of the current trit signature value by using the update rule $s\leftarrow
\lfloor s / 3\rfloor + t\cdot 3^{k-1}$, where $t$ is the next trit. 

We can replace the division with the fixed-point computation
$\bigl\lfloor\bigl(\bigl\lceil 2^{32}/3\bigr\rceil s\bigr) / 2^{32}\bigr\rfloor$
(this strategy works up to $k=19$ using $64$-bit integers), so by precomputing $\bigl\lceil 2^{32}/3\bigr\rceil$ and
$3^{k-1}$ the costly operations in the update of $s$ can be reduced to two
independent multiplications.

\subsection{Small counters}

The main implementation challenge, however, is that of reducing the
counter update area to improve the locality of access to the counters,
and possibly making it fit into some level of the processor cache.\footnote{When $k$ is large, this is not possible,
but we provide the option of improving memory access using large pages of the Translation Lookaside Buffer where available.} In a naive implementation,
we would need to use two ``large'' $64$-bit values to store the number of appearances of signature
$s$, and the sum of Hamming weights of the following words. Instead, we will use a
single ``small'' $32$-bit value, with a fourfold
space saving. In particular, we will use $13$ bits for the counter and $19$ bits for the
summation. This is a good choice as the largest Hamming weight for $w=32$ or $w=64$ is $64$, so if the
counter does not overflow, the summation will not, either.\footnote{With a similar argument, when $w=16$ one can choose $14$ bits for the counter and $18$ bits for the summation.}

We fix a \emph{batch size} and  update the small values blindly
through the batch. At the end of the batch, we update the large counters using
the current values of the small counters and zero the latter ones. At the same
time, we check that the sum of the small counters is equal to the batch size:
if not, a counter overflowed. Otherwise, we continue with the next batch,
possibly computing the transform and generating a $p$-value.

\subsection{Batch sizes}

How large should a batch be? We prefer larger batches, as access
to large counters will be minimized, but too large a batch will overflow small
counters. This question is 
interesting, as it is related to the \emph{mean passage time distribution} of the Markov chain having all possible signatures as states, and the probability of moving from signature $s$ to 
signature  $\lfloor s / 3\rfloor + t\cdot 3^{k-1}$ given by the probability of
observing the trit $t$. Let this probability be $p$ for the central values (trit $1$), and $(1-p)/2$ for the extremal values (trits $0$ and
$2$).
We are interested in the following question: given a $k$, $p$, and a batch size $B$, what is the
probability that a counter will overflow? This question can be reduced to the question: given $k$, $p$ and a batch size $B$,
what is the probability that the Markov chain after $B$ steps passes through the all-ones signature more than $2^{13}$ times?\footnote{It is obvious that the all-ones signature has
the highest probability in the steady-state distribution, and that by
bounding its probability of overflow we obtain a valid bound also for all other signatures.}
We want to keep this
probability very low (say, $10^{-100}$) so as not to interfere with the computation
of the $p$-values from the test; moreover, in this way, if we detect a counter
overflow we can simply report that we witnessed an event that cannot happen
with probability greater than $10^{-100}$, given that the source is random,
that is, a $p$-value.

Note that, in principle, we could use general results about Markov
chains~\cite[Theorem 7.4.2]{HunMTAP} which state that in the limit the number of
passages is normally distributed with mean and variance related to those of the
\emph{recurrence time distribution}, which can, in turn, be computed
symbolically using the Drazin inverse~\cite{HunVFPTMCAMT,MeyRGGITFMC}.

Since, however, no explicit bound is known for the convergence speed of the
limit above, we decided to compute exactly the mean passage time distribution
for the all-ones signature. To do this, we model the problem as a further Markov
chain with states $x_{c,j}$, where $0\leq c\leq b$, $b$ is a given overflow bound,
and $0\leq j< k$. 

The idea is that we will define transitions
so that after $u$ steps the probability of being in state $x_{c,j}$ will be the
probability that after examining $u$ $w$-bit values our current trit signature has a maximal suffix of $j$ trits equal to one,
and that we have counted exactly $c$ passages through the all-ones signature
($b$ or more, when $c=b$),
with the proviso
that the value $j=k-1$ represents both maximal suffixes of length $k-1$ and of length $k$ (we can
lump them together as receiving a one increases the passage count in both cases).
We use an initial probability distribution in which
all states with $c\neq0$ have probability zero, and all states with $c=0$ have
probability equal to the steady-state probability of $j$, which implies that we are
implicitly starting the original chain in the steady state. However, as argued also in~\cite{HunMTAP},
the initial distribution is essentially irrelevant in this context.

We now define the transitions so that the probability distribution of the new chain
evolves in parallel with the distribution of passage times of the original
chain (with the probability for more than $b$ passages lumped together):
\begin{itemize}
  \item all states $x_{c,j}$ have a transition with probability $1-p$ to $x_{c,0}$;
  \item all states $x_{c,j}$ with $j<k-1$ have a transition with probability $p$ to $x_{c,j+1}$;
  \item all states $x_{c,k-1}$ with $c<b$ have a transition with probability $p$ to $x_{c+1,k-1}$;
  \item there is a loop with probability $p$ on $x_{b,k-1}$. 
\end{itemize}
It is easy to show that after $u$ steps the sum of the
probabilities associated with the states $x_{b,-}$ is exactly the probability of
overflow of the counter associated with the all-ones signature.
We thus iterate the Markov chain (squaring the transition matrix is possible only for small $k$) until, 
say at step $B$, we obtain a probability of, say, $3^{-k}\bar p$: we can then guarantee that, given that
the source is random, running our test with batches of size $B$ we can observe overflow
only with probability at most $\bar p$.

This approach becomes unfeasible when we need to iterate the Markov chain
more than, say, $10^7$ times. However, at that point we use a very
good approximation: we apply a simple dynamic-programming scheme on the
results for $10^6$ steps to extend the results to a larger number of
steps. The idea is that if you know the probability $q_{u,c}$ that
the counter for the all-ones signature is $c$ after $u$ steps, then
approximately
\begin{align*}
q_{u+v,c} &= \sum_{f + g = c}q_{u,f}\cdot q_{v,g} \qquad\text{for $0\leq c <b$,}\\
q_{u+v,b} &= \sum_{f + g \geq b}q_{u,f}\cdot q_{v,g}.\\
\end{align*}
The approximation is due to the fact that the equations above implicitly
assume that the Markov chain is reset to its steady-state distribution
after $u$ steps, but experiments at smaller sizes show that the
error caused by this approximation is, as expected, negligible for large $u$. We thus
initialize $q_{u,c}$ for $u=10^6$ with exact data, and then we iterate
the process above to obtain the probabilities $q_{2^h u,c}$. These
probabilities are then combined in the same way to approximate
the probabilities associated with every multiple of $u$; at that 
point we can find the desired batch size by a binary search
governed by the condition that the probability associated with the
overflow bound $b$ is below a suitable threshold (e.g., $10^{-100}/3^k$).\footnote{It 
is worth noting that, based on the computations above, the normal approximation~\cite{HunMTAP}
is not very accurate even after a billion steps.}

In the end, we computed the ideal batch size as described above for $1\leq k\leq 19$ and 
included the result into our code (for example, when $w=64$ one obtains
$15\times 10^3$ for $k=1$, $2.3\times 10^6$ for $k=8$, and
$10^9$ for $k=16$). Combining all ideas described in this section, our test for
Hamming-weight dependencies with parameters $w=64$ and $k=8$ can analyze a 
terabyte of output of a $64$-bit generator in little more than 3
minutes on an Intel\textregistered{} Core\texttrademark{} i7-8700B CPU @3.20GHz. 
The $k=16$ test is an order of magnitude slower due to the larger memory
accessed. The Rust implementation of the test examines contiguous segments of the
output in parallel, with exactly the same results as a sequential run: on a 
$96$-core AMD EPYC 9R45 (Zen~5) CPU, the test with
$w=64$ and $k=8$ analyzes a terabyte of output of \xoroshirop[128] in about 3
seconds.

\subsection{Implementing the transform $T_k$}
\label{sec:transform}

In Figure~\ref{fig:transform} we show an in-place, recursive C implementation of the transform $T_k$
defined in Section~\ref{sec:hwd}.
The code is similar to analogous code for the Walsh--Hadamard transform or similar 
transforms based on Kronecker powers.

\begin{figure}
\small
\begin{verbatim}
void transform(double v[], int sig) {
    double * const p1 = v + sig, * const p2 = p1 + sig;

    for (int i = 0; i < sig; i++) {
        const double a = v[i], b = p1[i], c = p2[i];
        v[i] =  (a + b + c) / sqrt(3.0);
        p1[i] = (a - c) / sqrt(2.0);
        p2[i] = (2*b - a - c) / sqrt(6.0);
    }

   if (sig /= 3) {
       transform(v, sig);
       transform(p1, sig);
       transform(p2, sig);
   }
}
\end{verbatim}
\caption{\label{fig:transform}The code for the (in-place) transform described in
Section~\ref{sec:hwd}. It should be invoked with \texttt{sig} equal to
$3^{k-1}$.}
\end{figure}

The code assumes that the $3^k$-dimensional vector $\bm v$ is represented 
in the array \texttt{v}. The value associated with each signature is stored in a position equal to 
the signature (considered, as usual, as a base-$3$ numeral). In particular, the first $3^{k-1}$ values correspond
to signatures of the form $0s$, the following $3^{k-1}$ values
to signatures of the form $1s$, and the last  $3^{k-1}$ values to signatures of the form $2s$.
The function \texttt{transform()} must be invoked on \texttt{v} with the additional parameter \texttt{sig}
set to $3^{k-1}$.

We first note that if $k=1$ the function will just execute once the body of the for loop,
resulting in the in-place multiplication of the $3$-dimensional vector $\bm v$ by the base matrix $M$, as expected.

In the general case, the code scans the three subarrays \texttt{v}, \texttt{p1},
and \texttt{p2}, of length $3^{k-1}$, which as discussed above correspond to
signatures starting with $0$, $1$, and $2$, respectively.
With the notation of~(\ref{eq:rec}), these subarrays 
correspond to the subvectors $\bm v^0$, $\bm v^1$, and $\bm v^2$, respectively, and
it is immediate that the three subvectors appearing in the final result of~(\ref{eq:rec})
are computed in place by the for loop. After that computation, the
recursion applies by induction the transform with one dimension less to each of
the three subarrays in place. We conclude that \texttt{transform()} implements correctly in place the
transform $T_k$.

\section{Conclusions}

We have described a new test for Hamming-weight dependencies based on a unitary
transform. Properly implemented, the test is very powerful: for example, it
finds in a matter of hours bias in the dSFMT with $521$ bits of state and in
\xoroshirop[1024]; it can even find bias in WELL512a,
even though its transition matrix is much denser, and in the Tiny Mersenne Twister.
For these generators no bias was previously known beyond linearity tests.
In particular, the Hamming-weight tests in TestU01~\cite{LESTU01}, a
state-of-the-art testing framework, are unable to find any bias in several
generators of Table~\ref{tab:testhwd}, whereas all those generators fail our test.

Our test is very effective on $\F_2$-linear generators with relatively sparse
transition matrices, in particular when $wk$ is not smaller than the number
of bits of state of the generator.
In practice, the best results are obtained on generators with less than a few
thousand bits of state.

Similar to linearity tests, a failure in our test is an indication
of lesser randomness, but in general the impact will depend on the application.
We do not expect dependencies like those in~\xorshiftp[128] or the SFMT to be pernicious,
but they highlight a weakness of the associated $\F_2$-linear transformation~\cite{VigHTLGMT}.

\begin{acks}
The authors would like to thank Jeffrey Hunter for useful pointers to the
literature about mean passage times in Markov chains  
and Pierre L'Ecuyer for a number of suggestions that improved the
quality of the presentation.
\end{acks}

\bibliography{biblio}
\end{document}